\documentclass[%
 reprint,
 amsmath,amssymb,
 aps,
 prl
]{revtex4-2}

\usepackage{graphicx}% Include figure files
\usepackage{dcolumn}% Align table columns on decimal point
\usepackage{bm}% bold math
\usepackage{float}
\usepackage{xcolor}

\begin{document}

\preprint{APS/123-QED}

\title{Why Cold BGK Modes Are So Cool: Dispersion Relations from Orbit-Constrained Distribution Functions}% Force line breaks with \\

\author{Mikael Tacu}
 \email{mikael.tacu@cea.fr}
\affiliation{CEA, DAM, DIF F-91297 Arpajon, France}
\affiliation{Université Paris‐Saclay, CEA, Laboratoire
Matière en Conditions Extrêmes, F‐91680, Bruyères‐le‐Châtel, France}
\date{\today}
\begin{abstract}
We derive analytic dispersion relations for cold, orbitally constrained systems governed by the Vlasov equation. For magnetized plasmas, we obtain the first explicit relation for two-dimensional anisotropic BGK modes with finite magnetic field, showing that only a finite number of angular modes can become unstable and identifying a magnetic-field threshold for stabilization. In the gravitational case, we establish a bound on the growth rate of core perturbations, set by the potential’s curvature. These results clarify how orbital constraints shape the spectrum and growth of kinetic instabilities in cold, collisionless media.
\end{abstract}

\maketitle

Bernstein–Greene–Kruskal (BGK) modes~\cite{Bernstein1957} are exact, time-independent solutions of the Vlasov–Poisson equations, representing nonlinear phase-space structures that can persist in weakly collisional plasmas. In one-dimensional electrostatic settings, their properties and stability have been studied extensively~\cite{Schamel1972,Schwarzmeier1979} and are known to be associated with electron holes, solitary waves, and phase-space vortices~\cite{Hutchinson2017}. Observations in both laboratory and space plasmas~\cite{Ergun1998,Pottelette2005} indicate analogous structures in higher dimensions, but their existence and stability remain less understood. Ng and Bhattacharjee~\cite{Ng2005} constructed explicit two and three-dimensionsal BGK equilibria by including an angular momentum as a conserved quantity, enabling the existence of solutions in magnetized two dimensional and in unmagnetized three dimensional plasmas. Recent particle-in-cell simulations~\cite{McClung2024,Franciscovich2025} of such equilibria reported intriguing instability phenomena, including discrete angular mode patterns and a stability threshold depending on the background magnetic field. However, the physical mechanism of the spiral pattern observed in the azimuthal field, or of the stability threshold remains unexplained.

In this Letter, we present an exact linear stability analysis for a class of cold BGK equilibria constructed on orbit-constrained, Dirac delta-function distributions. For magnetized plasmas, we derive an explicit dispersion relation and show that only a finite number of angular modes can become unstable for any given potential. We identify a critical magnetic field above which all perturbations are linearly stable, and derive a local condition on the potential, guaranteeing core stability even for sub-threshold magnetic fields. These results clarify the mechanisms behind the discrete instability bands observed in previous simulations, and provide new analytical tools for exploring nonlinear phase-space dynamics.

We then apply the same method to the gravitational Vlasov–Poisson system, which governs the collisionless dynamics of globular clusters and galactic cores~\cite{binney2008}. Focusing on cold, orbitally constrained distributions relevant to the early evolution of such systems, we derive a strict upper bound on the growth rate of central perturbations, set entirely by the curvature of the gravitational potential. This analytic result connects central mass concentration to the timescale of kinetic redistribution, offering a new diagnostic for interpreting the early collapse phase seen in high-resolution N-body simulations~\cite{Vanalbada1982,labini2021}. The formalism is general and could be extended to other geometries or field configurations. Together, these results contribute to the theoretical understanding of phase-space instabilities in cold, long-range-interacting systems.

\textit{Plasma case. } 
We consider a collisionless plasma with a uniform ion density $n_i$ and a uniform background magnetic field $\textbf{B} = B_0\textbf{e}_z$, in cylindrical coordinates $(\textbf{e}_r,\textbf{e}_{\theta},\textbf{e}_z)$. It has been shown~\cite{Ng2005} that in the cylindrically symmetric case, with a radial electric field $\textbf{E} = -\nabla \psi$, any function of the Hamiltonian $w = v^2/2-\psi(r)$ and of the angular momentum $l = 2rv_{\theta}-r^2B_0$ is a solution to the following Vlasov equation (where $v = \sqrt{v_{\theta}^2+v_r^2}$ is normalized to the thermal velocity and $r$—to the Debye length), 
\begin{equation}
\begin{gathered}
\frac{\partial{f}}{\partial t}+ v_r\frac{\partial{f}}{\partial r} + \frac{v_{\theta}}{r} \frac{\partial f}{\partial \theta} + \left( \frac{v_{\theta}^2}{r} + \frac{d\psi}{dr} - v_{\theta}B_0 \right) \frac{\partial f}{\partial v_r} \\ - \left( \frac{v_r v_{\theta}}{r} + E_{\theta}- v_rB_0 \right) \frac{\partial f}{\partial v_{\theta}}= 0.
\end{gathered}
\end{equation}
Recently, some of these 2D BGK modes, corresponding to a particular distribution of the form $f(w,l) = (2\pi)^{-3/2} e^{-\omega}(1-he^{-kl^2})$, with $h\in (-\infty,1)$ and $k\in(0,\infty)$ constant parameters, have been shown via PIC simulations~\cite{McClung2024,Franciscovich2025} to exhibit a stability behavior dependent on a threshold value of the magnetic field $B_0$. To explore the stability in such a configuration, we consider the cold limit of distributions of the previous type. It writes as an exact solution to the Vlasov equation, on constrained orbits, and using Dirac delta functions as,
\begin{equation}
f_0(r,v_r,v_{\theta}) = g(r)\delta(v_r)\delta[v_{\theta}^2+r\psi'(r)-rv_{\theta}B_0].
\end{equation}

This idealized cold BGK mode is the simplest, yet physically relevant distribution, that allows a finite magnetic field $B_0$ and an arbitrary radial profile $g(r)$, where $g$ is any real positive function. This profile defines the potential $\psi(r)$ by Poisson equation, which writes, 
\begin{equation}
\frac{1}{r}\frac{d(r\psi'(r))}{d r} = g(r)\int_{\mathbb{R}} \delta[(v_{\theta} - v_{\theta}^{+})(v_{\theta} - v_{\theta}^{-})]\mathrm{d}v_{\theta} - 1,
\end{equation}
where, $2v_{\theta}^{\pm} = rB_0 \pm \sqrt{\Delta}$ are the roots of the polynomial $X^2-rB_0X+r\psi'(r)$, with $\Delta = r^2B_0^2-4r\psi'(r)$.

This cold distribution describes a plasma where electrons have zero radial velocity and where they turn, for any given radius $r$, at $v_{\theta} = v_\theta^{\pm}$ which depends on the magnetic field and on the potential. The electric field $E_r = -\psi'(r)$ is required to be positive, so that an equilibrium can be sustained even when $B_0 = 0$, in which case $\Delta \ge 0$. We then find then that $g$ is related to the potential by $d_r(r\psi'(r)) = 2rg(r)/\sqrt{\Delta}-r$.

By studying the linear stability of the previous distribution, we show, that for these functions, there exist a stability threshold based on the value of the magnetic field $B_0$. Let the time-dependent distribution be $f = f_0 + f_1$ and let's denote by $\dot{v}_r = v_{\theta}^2-rv_{\theta}B_0 + r\psi'(r)$, and $\dot{v}_{\theta} =v_rB_0-v_r v_{\theta}/r$ for more compact notations. The perturbed quantities are $f_1, E_r^1$ and $E_{\theta}^1$. The perturbed Vlasov equation becomes then,
\begin{equation}
\begin{gathered}
\frac{\partial{f_1}}{\partial t}+ v_r\frac{\partial{f_1}}{\partial r} + \frac{v_{\theta}}{r} \frac{\partial f_1}{\partial \theta} + \dot{v}_r\frac{\partial f_1}{\partial v_r} + \dot{v}_{\theta} \frac{\partial f_1}{\partial v_{\theta}} \\
= E_r^{1}g(r)\delta'(v_r)\delta(\dot{v}_r) + E_{\theta}^{1}g(r)(2v_{\theta}-rB_0)\delta(v_r)\delta'(\dot{v}_r).
\end{gathered}
\end{equation}
The stationary BGK mode has orbits in which $v_r = 0$. We suppose that after perturbation, electrons still follow these orbits. First, we seek for the simplest dispersion relation possible, second, small perturbations will remain confined on stable orbits for sufficiently small amplitudes and third, orbits with $v_r=0$ are still a valid solution of the Hamiltonian dynamics in the time dependent case. The perturbed distribution writes then $f_1 = \delta(v_r)F_1$, and the first moment as $F_1 = \int_{\mathbb{R}}f_1 \mathrm{d} v_r$. After integrating the previous equation in $v_r$, one gets $\partial_t F_1 + (v_{\theta}/r) \partial_{\theta}F_1 = E_{\theta}^{1}g(r)(2v_{\theta}-rB_0)\delta'(\dot{v}_r)$, since $\delta'(v_r)$ integrates to zero. This can be solved analytically by the method of characteristics, which gives : $F_1(r,\theta,v_{\theta},t) = g(r)(2v_{\theta}-rB_0)\delta'(\dot{v}_r) \int_{0}^{t}E_{\theta}\left( r,\theta-v_{\theta}(t-t_0)/r,t_0\right) \mathrm{d}t_0$. Here, we supposed that $F_1(r,\theta,v_{\theta},0) = 0$. For the perturbed Gauss equation, if we only retain the azimuthal field, then it writes $\partial_{\theta} E_{\theta}^{1} = -r\int_{\mathbb{R}} F_1 \mathrm{d}v_{\theta}$. A direct computation yields, 
\begin{equation}
\begin{gathered}
\frac{\partial E_{\theta}^1}{\partial \theta} = -\frac{g(r)}{\sqrt{\Delta}} \int_{0}^{t}  t_0\frac{\partial E_{\theta}^{1}}{\partial \theta}(r,\theta - v_{\theta}^{+}t_0/r,t-t_0)\mathrm{d} t_0 \\
 - \frac{g(r)}{\sqrt{\Delta}}\int_{0}^{t}  t_0\frac{\partial E_{\theta}^{1}}{\partial \theta}(r,\theta - v_{\theta}^{-}t_0/r,t-t_0)\mathrm{d} t_0.
\end{gathered}
\end{equation}
We can Fourier transform the periodic field $\theta \mapsto E_{\theta}^1(r,\theta,t)$, by writing $E_{\theta}^1 = \sum_{n=-\infty}^{+\infty} E_n(r,t) e^{in\theta} + c.c.$, which gives from the previous relation $\sqrt{\Delta}E_n(r,t) = - g(r)\int_{0}^{t} E_n(r,t-t_0)[e^{-inv_{\theta}^{+}t_0/r} + e^{-inv_{\theta}^{-}t_0/r}]\mathrm{d}t_0.$ We then perform a Laplace transform of each of these equations indexed by $n$, by writing $\hat{E}_n(r,\omega_n) = \int_{0}^{\infty} e^{i\omega_n t}E_n(r,t)\mathrm{d}t$. The dispersion relation follows then,
\begin{equation}
\sqrt{\Delta} = \frac{g(r)}{(\omega_n-nv_{\theta}^{+}/r)^2} + \frac{g(r)}{(\omega_n-nv_{\theta}^{-}/r)^2},
\end{equation}
where $\omega_n = \omega_{r,n} + i\gamma_n$ is the frequency of each mode indexed by $n$.

\textit{Discussion of the plasma case. }
To solve this dispersion relation, let us consider a given $\omega$ and introduce the following notations. Let $x_n = 2r(\omega_n - nB_0/2)/n\sqrt{\Delta}$ and $a_n = 4r^2g(r)/n^2\Delta^{3/2}$. Then, the frequency has to be chosen between the roots of the polynomial $P_n(X) = X^4-2(a_n+1)X^2+1-2a_n$. The roots of this polynomial are such that $x_n^2 = a_n+1 \pm \sqrt{a_n(a_n+4)}$. 
To study the stability of the modes, which are solutions to the dispersion relation, we use the Penrose criterion~\cite{Penrose1960}. The dispersion relation being given by a 4-th order polynomial, it is already an analytic function, and it has no poles in the upper half plane provided that $a_n\le 1/2$. So, the considered system is linearly stable when $a_n\le 1/2$ and unstable when $a_n>1/2$, with $\gamma_n>0$ in the latter case.

Let us first consider the stable case. The stability condition $a_n\le 1/2$ can also be written as $a_1\le 1/2$. Indeed, in that case $a_n\le a_1 \le 1/2$, and if $a_1>1/2$, then at least one mode is unstable, and so is the whole distribution. The latter stability condition can be written as $\psi'' + 2\psi'/r \le B_0^2/4-1$. A necessary condition directly following would be $B_0\ge2$, by taking the limit $r\to \infty$. However, this is not a sufficient condition, since we still have to impose the previous differential inequality, which can also be written as $[r^2 \psi'(r) - r^3(B_0^2/4-1)/3]' \le 0$. Integrating this is equivalent to using the Grönwall lemma~\cite{orauxens} with $B_0\ge2$ and amounts to no new information, since $E_r$ is positive. However, we already see that the condition is satisfied in the regions where $E_r$ is growing and also for $E_r \underset{r\to \infty}{\sim} E_r^{\infty}/r^2$, which is the case for a localized charge distribution by the Gauss theorem. We are left with the regions where $E_r$ decreases. If for example $E_r$ follows a $1/r^n$ law in some region, the stability condition becomes $n-2 \le (B_0^2/4-1)r^{n+1}$, which could be easily violated for a sufficiently large $n$ and for $r<1$. However, a too steep decrease of the electric field, and thus of the electron density (on Debye length scales), would be inconsistent with the assumption of a uniform charge distribution of the ion background. So, for realistic electric fields, with not too steep variations, the necessary and sufficient stability criteria could be still written as $B_0>2$.

Let us now discuss unconditional stability at small $r$. Since the global stability condition writes $\psi'' + 2\psi'/r \le B_0^2/4-1$, and with the assumption of $E_r(0) = 0$, we get by taking the limit $r\to 0$, that $\psi''(0) < B_0^2/12-1/3$. If this inequality is satisfied, then there exist a strictly positive $r_c = \inf \left\{ r > 0 \,\middle|\, \psi''(r) + 2\psi'(r)/r> 1 - B_0^2/4 \right\}$. Then, by definition, $\forall r\le r_c, a_1(r)<1/2$, meaning that the distribution is stable in that region, whatever the magnetic field $B_0$. Physically, this means that for a large class of potentials, a stable core region forms close to the axis regardless of the external magnetic field strength.

\begin{figure}[t]
\centering
\includegraphics[width=0.5\textwidth]{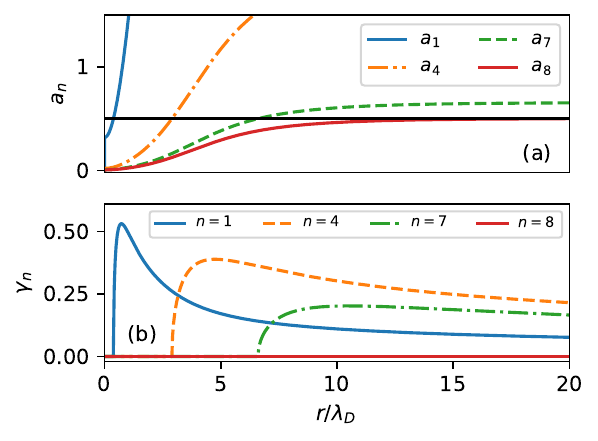}
\caption{In \textbf{(a)} we plot the stability-related parameters $a_n$, directly responsible for the stability of a given mode $n$. All the modes for which $a_n$ crosses the solid black line $y=1/2$ are unstable. In \textbf{(b)}, we have the corresponding growth rates $\gamma_n$. We see that while $a_n \le 1/2$, $\gamma_n = 0$. We also see that $a_8\le 1/2$. Since for all $n$, $a_{n+1}\le a_n$, we deduce the stability of all modes for $n\ge 8$. The results are plotted for a potential $\psi$ computed from the distribution $f(w,l) = (2\pi)^{-3/2} e^{-\omega}(1-he^{-kl^2})$. Here $B_0=0.25$, $k=0.4$ and $h = 0.9$.}
\label{fig1}
\end{figure}

For the unstable case, which is governed by the condition $a_n>1/2$, the growth rate can be written as $\gamma_n = \sqrt{\sqrt{a_n(a_n+4)}-a_n-1}/\sqrt{a_n}\Delta^{1/4}$ and the real frequency is given by $\omega_{r,n} = nB_0/2$. Since for a large enough $n$ we have $a_n<1/2$, it necessarily follows that in our case, \textbf{only a finite number of modes can grow unstable}. This is clearly represented on Fig.~\ref{fig1}\textbf{(a)} and \textbf{(b)}, where both $a_n$ and $\gamma_n$ are represented for a potential computed from Poisson equation for the distribution $f(w,l) = (2\pi)^{-3/2} e^{-\omega}(1-he^{-kl^2})$. To compute the potential, we numerically solved the differential equation (associated to the distribution by Poisson equation) $\psi''(r) = -\psi'(r)/r + e^{\psi(r)}(1-h e^{-kB_0^2r^4/4\gamma^2}/\gamma) - 1$, with $\psi'(0)=0$ and with $\psi(r) \underset{r \to \infty}{\to}0$.

\begin{figure}[t]
\centering
\includegraphics[width=0.5\textwidth]{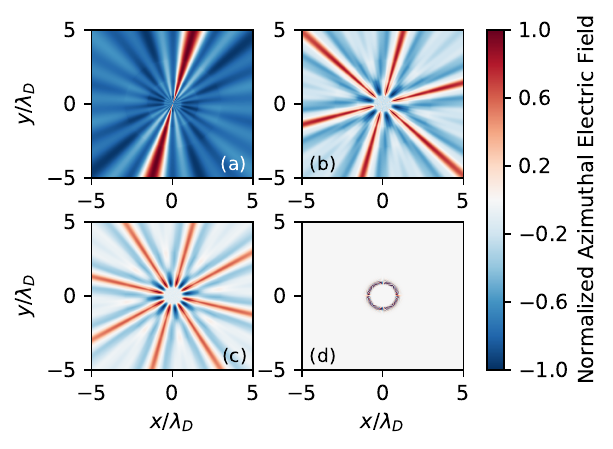}
\caption{Real part of the normalized azimuthal perturbed electric field in $(x,y)$ coordinates. The field is given by $E_{\theta}^1(r,\theta,t) = \sum_{n=1}^{n_{max}} E_n(r) e^{\gamma_n(r)t} \cos[n(\theta -B_0/2)t]$, where the $E_n$ were set equal to $E_n=1$ for clarity. The potential is the same as the one used in \cite{McClung2024, Franciscovich2025}, with $B_0=0.25$, $k=0.4$ and $h = 0.9$. Only eight modes were retained ($n_{max}=8$), since the rest are stable. In \textbf{(a)}, we have $t=2$, in \textbf{(b)}, $t=6$, in \textbf{(c)}, $t=10$ and in \textbf{(d)}, we have $t=100$. We see an unstable ring forming and persisting at large times.}
\label{fig2}
\end{figure}

Recent particle-in-cell (PIC) simulations qualitatively support our analytical predictions regarding finite-mode-number instabilities and magnetic-field stabilization thresholds. Specifically, McClung \textit{et al.} \cite{McClung2024} and later Franciscovich \textit{et al.}~\cite{Franciscovich2025} observed that cases with a magnetic field strength satisfying $B_0 \geq 2$ are stable, while instabilities occur for weaker fields, consistent with our derived stability threshold. The spiral pattern seen in Fig.~\ref{fig2} is also consistent with what was observed for the azimuthal field in PIC simulations performed previously. As shown on~\ref{fig2}\textbf{(a)}, \textbf{(b)} and \textbf{(c)}, the number of spiral arms may vary in time, which was also observed in Ref~\cite{McClung2024}. In the linear case however, as seen on Fig.~\ref{fig2}\textbf{(d)}, the field will eventually be dominated by the region surrounding the maximum value of $a_1$, consistent with the growth rates represented on Fig.~\ref{fig1}, which also delimits the unconditionally stable region near the core.

\textit{Gravitational case. } In the gravitational analog, we consider a collection of stars evolving in a spherically symmetric, self consistent potential $\psi(r)$, which is general in this study, but could be for example a Plummer type potential $\psi(r) = -GM/\sqrt{r^2+a^2}$, shown to emphasize the sign difference with the plasma case. Here $G$ is the universal gravitational constant, $M$ is the object's mass and $a$ is the Plummer radius. The Vlasov equation in 1D2V geometry writes as \cite{binney2008}, 
\begin{equation}
\frac{\partial f}{\partial t} + v_r\frac{\partial f}{\partial r} + \left( \frac{v_{\perp}^2}{r} - \frac{d\psi}{dr}\right) \frac{\partial f}{\partial v_r} - \frac{v_r v_{\perp}}{r} \frac{\partial f}{\partial v_{\perp}} = 0,
\label{1d2v}
\end{equation}
where $v_{\perp} = \sqrt{v_{\theta}^2 + v_{\varphi}^2}$. We modify the original construction by Ng \textit{et al.}, by using the fact that $v_{\perp} \ge 0$. Instead of evaluating the Dirac distribution in $v_{\perp}^2/r-\psi_0'(r)$, we evaluate it in $v_{\perp} - \sqrt{r\psi_0'(r)}$, where $\psi_0$ is the stationary potential. This makes the distribution simpler and more suitable for a perturbative study. It writes,
\begin{equation}
f_0(r,v_r,v_{\perp}) = g(r)\delta(v_r)\delta\left[v_{\perp} - \sqrt{r\psi_0'(r)}\right].
\end{equation}
Physically, we suppose that the cluster is initially cold and that all the stars follow circular orbits with $v_{\perp} = \sqrt{r\psi'(r)}$. The main reason to consider such a distribution is mathematical. It is the simplest distribution, solution to Vlasov equation, that supports a generic potential $\psi_0$. In the gravitational case Poisson's equation writes,
\begin{equation}
\frac{1}{r^2} \frac{\partial}{\partial r}\left(r^2\frac{\partial \psi}{\partial r}\right) = 8\pi^2 G \int_{\mathbb{R}}\mathrm{d} v_r \int_{0}^{+\infty} v_{\perp}\mathrm{d} v_{\perp} f(r,v_r,v_{\perp},t).
\end{equation}
For the stationary potential, we have the following relation: $8\pi^2Gr\sqrt{r\psi_0'(r)}g(r) = 2\psi_0'(r) + r\psi_0''(r)$. Now, to study the stability of this class of distributions, we suppose that in the linear phase, the stars continue to follow circular orbits in the stationary potential $\psi_0$. This simplifies the treatment of the problem while keeping it physically insightful. Supposing instead that the stars continue on $v_r=0$ orbits, amounts to no dispersion relation and to inconditional stability. We write thus $f(r,v_r,v_{\perp},t) = f_0(r,v_r,v_{\perp}) + F_1(r,v_r,t)\delta(v_{\perp}-\sqrt{r\psi_0'(r)})$ and after linearizing Eq.(7) and integrating in $v_{\perp}$, we obtain, 
\begin{equation}
\frac{\partial F_1}{\partial t} + v_r \frac{\partial F_1}{\partial r} + \frac{v_r}{r}F_1 = \frac{\partial \psi_1}{\partial r}g(r)\delta'(v_r),
\end{equation}
where $\psi(r,t) = \psi_0(r)+\psi_1(r,t)$. This equation, like the previous one in the plasma case, can be integrated on the characteristics, by noticing that if $x = r+v_r t$, then  $d_t[xF_1(x,v_r,t)] = x\partial_r \psi_1(x,t)g(x)\delta'(v_r)$. After integration we obtain with $r' = r-v_r(t-t')$ that $rF_1(r,v_r,t) = \int_{0}^{t} r'\partial_r\psi_1(r',t')g(r')\delta'(v_r)\mathrm{d}t'$. Further, we use the fact that $q(s)\delta'(s) = -q'(0)\delta(s) + q(0)\delta'(s)$ for any function $q$. By the way, the obtained expression for $F_1$ is of the form $g_0(r,t)\delta(v_r)+h_0(r,t)\delta'(v_r)$. The $\delta'(v_r)$ term will vanish when reported in the density. However, it gives an interesting idea as an ansatz for searching solutions to the full 1D2V Vlasov equation. One could look to solutions of the form $g_1(r,v_\perp,t)\delta(v_r)+h_1(r,v_\perp,t)\delta'(v_r)$, then take the moments. The $\delta'(v_r)$ term does not contribute to Poisson equation, but it ensures a mathematical coherence, since a Dirac delta derivative $\delta'(v_r)$ appears naturally when propagating the unperturbed distribution, coming from the term $\partial_{v_r}f_0$.

We Laplace transform the linearized Poisson equation and introduce for that \textbf{$\tilde{\psi}_1(r,\omega) = \int_{0}^{\infty} e^{i\omega t} \psi_1(r,t) \mathrm{d}t$}. To express the dispersion relation, let us denote by $z = \partial_r \tilde{\psi}_1$ and by $z'= \partial_rz$. Then after some straightforward manipulations, we obtain,
\begin{equation}
\left(r + \frac{2\psi'_0 + r\psi''_0}{\omega^2} \right)z' + \left( 2 - \left[1+\frac{rg'}{g}\right]\frac{2\psi'_0 + r\psi''_0}{r\omega^2}\right)z = 0.
\end{equation}
This differential equation is of the form $c_1(r)z'(r)+c_0(r)z(r)=0$ and its solution integrated from an $r_0>0$ can be written as $z(r) = z(r_0)e^{-\int_{r_0}^{r}c_0(r_1)/c_1(r_1)\mathrm{d}r_1}$. Since, $c_0(r)/c_1(r) \underset{r\to\infty}{\sim} 2/r$, we see after integrating again in $r$, that $\psi_1 \underset{r\to\infty}{\sim} -z(r_0)/r$, which is a consistent behaviour at infinity for a gravitational potential. For $r=0$, we require a finite value of the perturbed potential with $z'(0) = 0$. However, since $c_0(r)/c_1(r)\underset{r\to 0}{\sim}2/r(1+3\psi''(0)/\omega^2)$, the only way this could be satisfied in the limit $r_0 \to 0$ is for $1+3\psi''(0)/\omega^2$ to be strictly negative (or zero, which will require to go to the second order in series expansion of $c_1$). Since $\psi_0''(0)>0$ in most of the cases (for a Plummer potential for example, $\psi_0''(0) = GM/a^3$), it follows that the frequency $\omega^2$ is negative. Let's denote it by $\omega = i\gamma$, where $\gamma$ is the growth rate. The considered class of distributions is unstable, for the growth of perturbations such that, 
\begin{equation}
\gamma \le \sqrt{3\psi''(0)}.
\end{equation}
\textit{Discussion of the gravitational case. }The instability condition derived from the orbit-constrained distribution imposes an upper bound on the growth rate of perturbations near the cluster core. This constraint arises from requiring a finite, regular gravitational potential at the origin. Physically, this means that there can not be an arbitrary fast redistribution in a cold, collisionless core, without a correspondingly large central potential curvature. This limits the development of kinetic instabilities. Since $\psi_0''(0)$ depends on the central mass density, this links the dynamical timescale of instability growth to the underlying potential structure. For instance, in a Plummer potential, we have $\psi_0''(0) = GM/a^3$, yielding $\gamma \lesssim \sqrt{3GM/a^3}$, which is on the order of the inverse dynamical time. This is different from the classical Jeans instability, which by the way, is suppressed in this case, along with the radial orbit instability, because of the infinitely small dispersion in radial velocity. This theoretical bound offers a diagnostic: in principle, it implies that if one observes a fast redistribution rate in a cold, collisionless core, whether in simulations or inferred from transient structures, the local curvature of the potential must be correspondingly steep. Conversely, a shallow potential, such as in a globular cluster without a central black hole, imposes a slower maximum growth. The presence of an intermediate-mass black hole~\cite{Haberle2024}, by contrast, steepens $\psi''(0)$, thereby allowing faster instability-driven evolution. This suggests that in systems dominated by central mass concentrations, kinetic mechanisms may drive fast early evolution even before collisional processes become important. This could allow for interpreting high-resolution N-body simulations and potentially constraining unseen central masses via dynamical arguments alone. 

As for cold, collisionless stellar systems, numerical simulations~\cite{Vanalbada1982} have shown that early core evolution is shaped by orbitally constrained dynamics. In the case of initially cold and spherically symmetric configurations~\cite{merritt1985}, their violent relaxation leads to central cusps and quasi-stationary cores. Recent high-resolution studies~\cite{labini2021,rozier2019} confirm that during this early phase, the central region is dominated by coherent, near-circular orbits—consistent with our analytic assumptions. In this regime, collisional effects are negligible and phase-space instabilities govern the redistribution process. Our derived bound on the instability growth rate provides a kinetic limit to this redistribution, directly linking it to the curvature of the gravitational potential. This echoes findings in simulations with and without central black holes~\cite{breen2013}, where deeper central potentials accelerate core evolution. Although these simulations do not directly impose our cold, orbit-constrained conditions, they suggest that the mechanism captured by our analytic model may naturally emerge in the early stages of globular cluster evolution, before relaxation isotropizes the velocity distribution.

\textit{Relevance of cold BGK distributions.} While the exact delta-function distribution employed in our analytical model represents an idealized limit, practical numerical implementations necessarily involve finite but small velocity dispersions due to numerical constraints. This could be tested in principle by Vlasov simulations.  However, as recently shown~\cite{tacu2024}, one has to be careful with discontinuous distributions which have sharply varying derivatives, while performing Vlasov simulations, let alone with Dirac delta-functions as an initial condition. This is especially true with the commonly used semi-Lagrangian schemes. A more direct method of testing the exact findings in this paper would be a test-particle algorithm, where the conservation in phase space would be used to propagate the initially cold distribution, and then Poisson equation would be solved numerically. This is left for future work, since a careful treatment of the radial Dirac term $\delta(r-r_{orb})$ would be necessary in the Poisson's equation, where $r_{orb}$ would follow the time-dependent characteristics of Eq.~\ref{1d2v}.

\textit{Summary. } In summary, we have developed a unified kinetic framework for deriving analytic dispersion relations in cold, orbitally constrained systems governed by the Vlasov equation. In the plasma context, this approach yields the first explicit dispersion relation for cold, anisotropic BGK modes in two dimensions with finite magnetic field, containing a finite set of unstable angular modes and a threshold for magnetic stabilization. Applied to self-gravitating systems, the same method establishes a strict upper bound on the growth rate of central perturbations, determined by the curvature of the gravitational potential. This result implies that cold, collisionless cores cannot undergo rapid redistribution without significant central mass concentration—linking the dynamics of instability growth to core structure. Together, these findings highlight a shared kinetic mechanism in long-range-interacting systems. The formalism is broadly applicable and may enable new dispersion relations in other geometries. Future analytic work, simulations, and observational studies informed by these results could advance our understanding of phase-space instabilities across both plasma and astrophysical settings.

\begin{acknowledgments}
The author would like to thank Charles Ruyer, Vanina Recoules and Jean-Christophe Pain for carefully reading the manuscript. Useful discussions with Didier Bénisti, Serge Bouquet, Matthias Pautard and Robin Piron are gratefully acknowledged.
\end{acknowledgments}

\end{document}